\documentclass[aps,prB,reprint,superscriptaddress,twocolumn]{revtex4-1}
\usepackage{amsmath,amssymb}
\usepackage{graphicx}
\usepackage{color}
\usepackage{ulem}
\draft

\begin{document}
\makeatletter 
\makeatother

\title{An ambipolar single-charge pump in silicon}
\author{Gento Yamahata}
\email{E-mail: gento.yamahata@ntt.com}
\affiliation{NTT Basic Research Laboratories, NTT Corporation, 3-1 Morinosato Wakamiya, Atsugi, Kanagawa 243-0198, Japan}

\author{Akira Fujiwara}
\affiliation{NTT Basic Research Laboratories, NTT Corporation, 3-1 Morinosato Wakamiya, Atsugi, Kanagawa 243-0198, Japan}
\begin{abstract}
The mechanism of single-charge pumping using a dynamic quantum dot needs to be precisely understood for high-accuracy and universal operation toward applications to quantum current standards and quantum information devices. The type of charge carrier (electron or hole) is an important factor for determining the pumping accuracy, but it has been so far compared just using different devices that could have different potential landscapes. Here, we report measurements of a silicon ambipolar single-charge pump. It allows a comparison between the single-electron and single-hole pumps that share the entrance tunnel barrier, which is a critical part of the pumping operation. By changing the frequency and temperature, we reveal that the entrance barrier has a better energy selectivity in the single-hole pumping, leading to a pumping error rate better than that in the single-electron pumping up to 400 MHz. This result implies that the heavy effective mass of holes is related to the superior characteristics in the single-hole pumping, which would be an important finding for stably realizing accurate single-charge pumping operation.
\end{abstract}

\maketitle
Ambipolar devices allow the physical properties of electrons and holes to be directly compared. To clarify how different types of charge carriers affect device operation, ambipolar quantum dots (QDs) in various materials have been intensively studied focusing on DC transport measurements \cite{10.1063/1.123464,PJ-CNT, 10.1063/1.2207494, PhysRevLett.103.046810,10.1063/1.4898704,doi:10.1021/acs.nanolett.5b01706, 10.1063/1.5048097, doi:10.1021/acs.nanolett.0c03227} and charge sensing \cite{ PhysRevB.101.201301, 10.1063/5.0040259, doi:10.1021/acs.nanolett.2c04417}. A dynamic QD with a tunable barrier has also been used for a single-charge pump, which accurately transfers single charges by a clock control \cite{UnivRev}. There are various applications of single-charge pumps such as quantum current standards \cite{pekola-rev}, quantum information devices \cite{SEsouce1,GYnnano, Jcol, Ncol,SAWcol}, single-photon sources \cite{SAWsp,norimoto2024photonemissionhotelectron}, and quantum sensing \cite{Nathan_samp}. While many high-accuracy measurements of single-electron pumping have been reported \cite{gib1,PTB-ulca1, Kriss-HR, NPL-NTT1, Zhao_pump,RobPTB, KrissAIST, LHeSi,Si2GHz,doi:10.1021/acs.nanolett.3c02858}, the pumping accuracy needs to be further understood and improved to reproducibly achieve operation with an accuracy of better than 0.1 ppm \cite{Fujiwara_2023}. One idea is single-hole pumping, in which the tunnel barrier that forms the entrance of the QD should be highly energy selective due to the heavy effective mass \cite{holeJAP}. However, the reported single-hole pumps were only unipolar ones \cite{holeJAP,myaplhole,RosHole}. Therefore, a single-charge pump with an ambipolar QD is an important next step in comparing and understanding the difference between electrons and holes in terms of pumping accuracy.

In this letter, we report an ambipolar single-charge pumping achieved using a device with n- and p-type contacts. It enables us to compare the single-electron pumping and single-hole pumping using the same entrance tunnel barrier. We investigate frequency and temperature dependence of the pump characteristics and show better characteristics in the case of the single-hole pumping up to 400-MHz operation, implying the importance of the heavy effective mass of holes for high-accuracy operation.

\begin{figure}
\begin{center}
\includegraphics[pagebox=artbox]{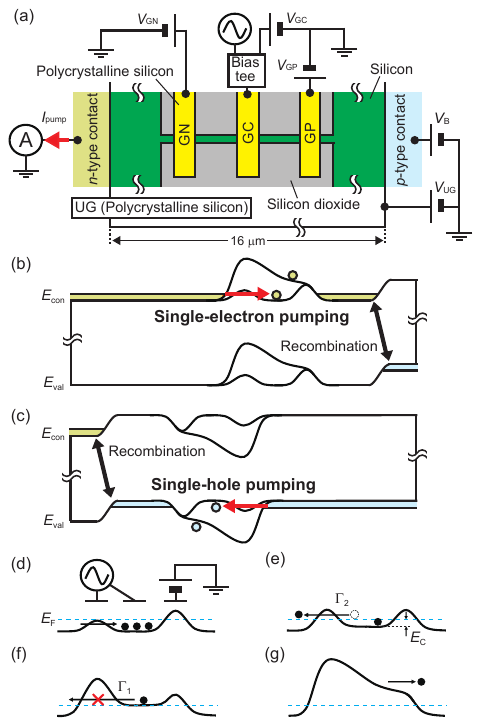}
\end{center}
\caption{(a) Schematic of the device structure with electrical connections. DC voltages ($V_{\mathrm{GN}}, V_{\mathrm{GC}}, V_{\mathrm{GP}}, V_{\mathrm{B}}, V_{\mathrm{UG}}$) are supplied by DC voltage sources (Yokogawa GS200). An arbitrary waveform generator (Keysight M8195A) generates a sinusoidal voltage $A_{\mathrm{amp}}\mathrm{sin}(2 \pi f_{x} t)$, where $A_{\mathrm{amp}}$ is the amplitude (50 $\Omega$ output impedance) and $f_{x}$ is the frequency with $x$ MHz. This sinusoidal signal is amplified by a 15-GHz low-noise amplifier (Tektronix PSPL8003) followed by a 3-dB attenuation. $I_{\mathrm{pump}}$ is converted into voltage by a current amplifier (NF CA5351). The converted voltage is measured by a voltmeter (Keysight 3458A). (b)(c) Energy band diagrams in the case of the single-electron pumping and single-hole pumping, respectively. $E_{\mathrm{con}}$ and $E_{\mathrm{val}}$ are the conduction band bottom and valence band top, respectively. (d)-(g) Potential diagrams for explaining the mechanism of the single-charge pumping using a dynamic QD, where $E_{\mathrm{F}}$ is the Fermi level, $E_{\mathrm{C}}$ is the charging energy of the QD, and $\Gamma_{n}$ is the escape rate from the QD with $n$ charges. The black dot is a charge carrier.}
\label{f1}
 \end{figure}

The device has a double-layer gate structure [an upper gate (UG) and three lower gates (GN, GC, GP)] on a silicon nanowire [Fig.~\ref{f1}(a)]. The fabrication process is as follows. We used a silicon-on-insulator wafer with a buried oxide thickness of 400 nm. The silicon nanowire was formed using electron-beam lithography and dry etching. Next, a gate oxide was grown by thermal oxidation. After growth of n-type polycrystalline silicon using chemical vapor deposition, the lower gates were formed using electron-beam lithography and dry etching. Then, the interlayer oxide was grown by thermal oxidation. After n-type polycrystalline silicon was grown by chemical vapor deposition, UG was formed using optical lithography and dry etching. The n- and p-type contact regions outside UG were formed by implanting phosphorous and boron, respectively, with a resist as a mask of the implantation. Since the silicon nanowire is far from these regions, it is not doped with impurities. Finally, aluminum ohmic contacts were formed by vacuum deposition. The thickness and width of the silicon nanowire are 15 and 20 nm, respectively. The gate length and space between the adjacent lower gates are 10 and 80 nm, respectively. The gate oxide thickness under the lower gate is 30 nm.

A DC forward bias $V_{\mathrm{B}}$ of about 1 V was applied to the p-type contact to recombine electrons and holes near the p-type [Fig.~\ref{f1}(b)] or n-type [Fig.~\ref{f1}(c)] contacts, allowing conduction between the n- and p-type contacts. Electrons or holes in the silicon nanowire and adjacent regions can be induced by applying sufficiently positive or negative DC gate voltage $V_{\mathrm{UG}}$ to UG, respectively. We applied DC voltage $V_{\mathrm{GC}}$  combined with a sinusoidal voltage with frequency $f_{x}$ to GC, where $x$ is the value of the frequency in MHz. Negative or positive $V_{\mathrm{GC}}$'s are necessary to form an entrance barrier for the single-electron pumping or single-hole pumping, respectively. In addition, we applied DC negative voltage $V_{\mathrm{GP}}$ to GP and DC positive voltage $V_{\mathrm{GN}}$ to GN to form an exit barrier. Depending on the charge carrier type in the silicon nanowire, we can observe single-electron pumping from GC to GP [Fig.~\ref{f1}(b)] and single-hole pumping from GC to GN [Fig.~\ref{f1}(c)]. The current $I_{\mathrm{pump}}$ was measured at the n-type contacts. The measurements were performed using a dilution refrigerator with a temperature control ($T=50$ mK $\sim 16$ K). 

We pumped single charges using the tunable-barrier dynamic QD formed between the adjacent two lower gates [Figs.~\ref{f1}(d)-\ref{f1}(g)] \cite{tunable-barrier1}. When the entrance barrier is sufficiently low, some charges are loaded into the QD from the charge source on the entrance barrier side [Fig.~\ref{f1}(d)]. After that, both entrance barrier and QD rise because of a capacitive coupling between the entrance gate and QD. When the electrochemical potential of the QD is higher than the Fermi level $E_{\mathrm{F}}$, the stored charges escape back to the charge source on the entrance barrier side [Fig.~\ref{f1}(e)]. We define rate $\Gamma _{n}$ for the charge escape from the QD with $n$ charges. Since $\Gamma _{n}$ becomes exponentially small during the rise of the entrance barrier, a single charge can be captured when $\Gamma _{n}$ is much lower than the characteristic rate $\Gamma _{\mathrm{inc}}$ of the escape \cite{GYgval}. The capture accuracy of a single charge becomes better when $\Gamma_{2}/\Gamma_{1}$ is larger. Note that charging energy $E_{\mathrm{C}}$ [Fig.~\ref{f1}(e)] and the time difference between Figs.~\ref{f1}(e) and \ref{f1}(f) are important factors for $\Gamma_{2}/\Gamma_{1}$. The captured single charge is eventually ejected to the opposite side over the exit barrier [Fig.~\ref{f1}(g)]. When a single charge is transferred each cycle, the current level is $I_{\mathrm{pump}}=ef_{x}$, where $e$ is the elementary charge. 

\begin{figure}
\begin{center}
\includegraphics[pagebox=artbox]{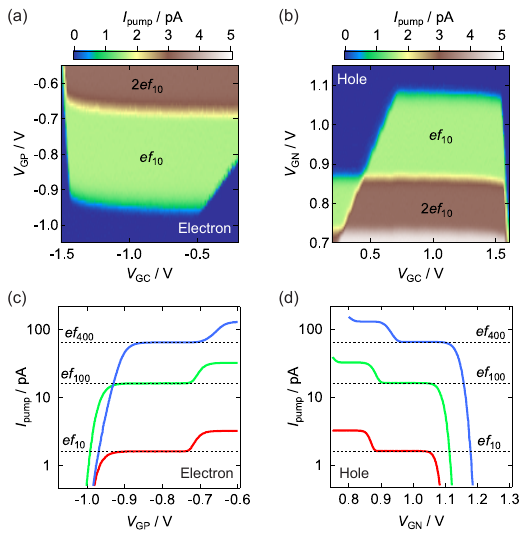}
\end{center}
\caption{(a) $I_{\mathrm{pump}}$ generated by the single-electron pumping as a function of $V_{\mathrm{GC}}$ and $V_{\mathrm{GP}}$ at $f_{10}$ and $T = 6.8$ K, where $A_{\mathrm{amp}}=0.175$ V (nominally, 1.4 V at GC), $V_{\mathrm{UG}}= 0.5$ V, $V_{\mathrm{GN}}=1$ V, and $V_{\mathrm{B}}=1$ V. (b) $I_{\mathrm{pump}}$ generated by the single-hole pumping as a function of $V_{\mathrm{GC}}$ and $V_{\mathrm{GN}}$ at $f_{10}$ and $T = 6.8$ K, where $A_{\mathrm{amp}}=0.175$ V, $V_{\mathrm{UG}}= -1$ V, $V_{\mathrm{GP}}=-1$ V, and $V_{\mathrm{B}}=1.05$ V. (c) $I_{\mathrm{pump}}$ generated by the single-electron pumping as a function of $V_{\mathrm{GP}}$ at $f_{10}$, $f_{100}$, and $f_{400}$, where $A_{\mathrm{amp}}=0.175$ V, $V_{\mathrm{UG}}= 0.5$ V, $V_{\mathrm{GN}}=1$ V, and $V_{\mathrm{B}}=1$ V. For the 10- and 100-MHz operations, $V_{\mathrm{GC}}=-1$ V and $T=1.2$ K. For the 400-MHz operation, $V_{\mathrm{GC}}=-1.05$ V and $T=1$ K. These differences are not essentially important. (d) $I_{\mathrm{pump}}$ generated by the single-hole pumping as a function of $V_{\mathrm{GN}}$ at $f_{10}$, $f_{100}$, and $f_{400}$, where $A_{\mathrm{amp}}=0.175$ V, $V_{\mathrm{UG}}= -1$ V, $V_{\mathrm{GP}}=-1$ V, and $V_{\mathrm{B}}=1.05$ V. For the 10- and 100-MHz operations, $V_{\mathrm{GC}}=1.1$ V and $T=1.2$ K. For the 400-MHz operation, $V_{\mathrm{GC}}=1.2$ V and $T=1$ K. These differences are not essentially important. }
\label{f2}
 \end{figure}

Figures~\ref{f2}(a) and \ref{f2}(b) show typical pumping current maps for the single-electron pumping and single-hole pumping, respectively. The horizontal and vertical axes are the entrance and exit gate voltages, respectively. Both types of single charges were successfully pumped using the same entrance gate (GC). The observation of the pumping current indicates that the pumped charges always recombine near the contact regions. Figures~\ref{f2}(c) and \ref{f2}(d) show frequency dependence of the single-electron pumping and single-hole pumping current as a function of the exit gate voltage, respectively. We focus on these three frequencies (10, 100, 400 MHz) in this letter.

\begin{figure}
\begin{center}
\includegraphics[pagebox=artbox]{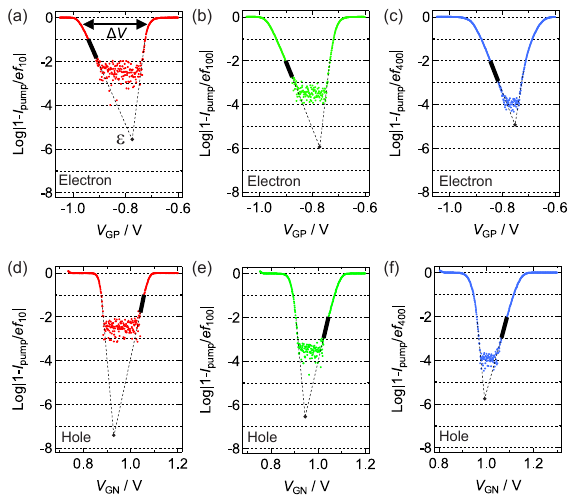}
\end{center}
\caption{(a)-(f) $\mathrm{Log}|1-I_{\mathrm{pump}}/ef_{x}|$ as a function of $V_{\mathrm{GP}}$ (a,b,c) or $V_{\mathrm{GN}}$ (e,f,g). The measurement conditions are the same as those in Figs.~\ref{f2}(c) and \ref{f2}(d). The black lines are linear fits to the data. The dashed lines are linear extensions of the data. The black dots are the intersection of the extended lines. $\Delta V$ is the plateau width.}
\label{f3}
 \end{figure}

To compare single-electron pumping and single-hole pumping in more detail, it is useful to investigate a deviation of $I_{\mathrm{pump}}$ from a quantized value of $ef_{x}$. We took the logarithm of the deviation normalized by $ef_{x}$ at 10, 100, and 400 MHz and plotted it [Figs.~\ref{f3}(a)-\ref{f3}(f)]. Since our measurements were performed using commercially available equipment, we can only measure the deviations on the order of $10^{-4}$ at best, which becomes better as the current level increases, i.e., $f_{x}$. A theoretical lower bound $\epsilon$ of the error rate can be defined as the intersection of straight lines extending the left and right lines of the deviation plot \cite{GYgval,LHeSi,holeJAP}. Note that this estimation would be reasonable even at the level of 0.1 ppm \cite{Si2GHz}. As seen in Figs.~\ref{f3}(a)-\ref{f3}(f), $\epsilon$ in the single-hole pumping is lower than that in the single-electron pumping in these frequency regimes.

To more quantitatively understand these results, we extracted slope $A$ of the deviation at the rise of the current to $ef_{x}$. Since the slope of $|1-I_{\mathrm{pump}}/ef_{x}|$ is theoretically exponential in the regime with a sufficiently low error rate (i.e., sufficiently large $V_{\mathrm{GP}}$ or small $V_{\mathrm{GN}}$) \cite{holeJAP}, we performed a fit by $-A(V_{\mathrm{GP}}-V_{0})$ or $A(V_{\mathrm{GN}}-V_{0})$ to $\mathrm{Log}|1-I_{\mathrm{pump}}/ef_{x}|$ to extract $A$. Because of the limitation of the measurement accuracy, we selected the range of the fit between -1 to -2 for the 10-MHz data and between -2 to -3 for the 100- and 400-MHz data (black lines in Fig.~\ref{f3}). In addition, we extracted the plateau width $\Delta V$ between $I_{\mathrm{pump}}/ef_{x} = 0.5$ and $1.5$ indicated by the black arrow in Fig.~\ref{f3}(a).

\begin{figure}
\begin{center}
\includegraphics[pagebox=artbox]{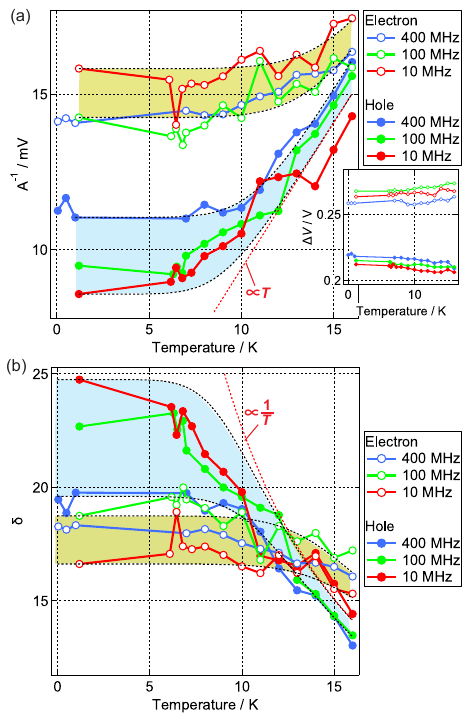}
\end{center}
\caption{(a) $A^{-1}$ extracted from the linear fit shown in Fig.~\ref{f3} as a function of temperature at 10, 100, and 400 MHz for the single-electron pumping and single-hole pumping. The dashed curves that form the edge of the light yellow and light blue eye guides are phenomenological fits to the 10- and 100-MHz data of the single-electron pumping and to the 10- and 400-MHz data of the single-hole pumping with a function of $A_{0}^{-1}(T_{0}^8+T^{8})^{1/8}/T_{0}$, where $A_{0}^{-1}$ is equal to $A^{-1}$ at around 1 K. We obtained roughly estimated values of $T_{0}$ from these fits. The inset shows $\Delta V$ extracted from the data shown in Fig.~\ref{f3} as a function of temperature. Each data point corresponds to that of the main panel. (b) $\delta$ extracted from $A\Delta V$ in Fig.~\ref{f4}(a) as a function of temperature at 10, 100, and 400 MHz for the single-electron pumping and single-hole pumping. The dashed curves that form the edge of the light yellow and light blue eye guides are $\delta _{0}T_{0}/(T_{0}^8+T^{8})^{1/8}$, where $\delta _{0}$ is $\delta$ at around 1 K and $T_{0}$ is a value extracted from the fits in (a).}
\label{f4}
 \end{figure}

Figure~\ref{f4}(a) shows $A^{-1}$ as a function of temperature for the single-electron pumping and single-hole pumping at 10, 100, and 400 MHz. Theoretically, $A^{-1}$ is proportional to temperature in the thermal hopping regime at high temperature and to characteristic temperature $T_{0}$ in the tunneling regime at low temperature, where $T_{0}=\hbar \sqrt{S}/(2 \pi k \sqrt{m})$, $\hbar$ is the reduced Planck constant, $k$ is the Boltzmann constant, $m$ is the effective mass of the charge carrier, and $S$ is the curvature of the entrance barrier assuming parabolic \cite{holeJAP}. Since we use the same entrance barrier, we assume that $S$ is the same for the electron and hole. When $T_{0}$ is low, the difference in the tunnel rate between different barrier heights relative to the QD energy becomes large, meaning a better energy selectivity for the entrance barrier. The light yellow and light blue eye guides are illustrated on the basis of the transition from the tunneling to thermal hopping regimes. Here, we simply assume a phenomenologically smooth transition between the thermal hopping and tunneling using a function of $(T^{8} + T_{0}^{8})^{1/8}$. An accurate discussion of this transition deserves further investigation. $A^{-1}$ for the single-hole pumping is proportional to temperature above about 11 K, while the $A^{-1}$ for the single-electron pumping does not reach the region proportional to temperature. A rough estimation is that $T_{0}$ for the single-electron pumping and single-hole pumping are 9-11 K and 14-15 K, respectively, which are similar values to those in the previous reports about our silicon pumps \cite{holeJAP, Nathan_cur}. For a crude theoretical evaluation of the difference in $T_{0}$, we use the effective masses of electrons ($m=0.19m_{0}$) and holes ($m=0.49m_{0}$) in silicon \cite{Sze2}, where $m_{0}$ is the free electron mass. Based on the expression of $T_{0}$ introduced above, $T_{0}$ for electrons is $\sqrt{0.49/0.19}\sim 1.6$ times larger than $T_{0}$ for holes. The order of magnitude of the difference in $T_{0}$ estimated from the experiments is similar to that estimated from the theoretical consideration. This implies that the observed difference in $T_{0}$ may be due to the heavy effective mass of holes. Note that the frequency dependence of $T_{0}$ in the case of the single-hole pumping is slightly larger than that in the case of the single-electron pumping. Although the reason is not clear, we speculate that since the entrance barrier height should be lower in the case of the single-hole pumping due to the heavy effective mass, single-hole pumping may be sensitive to the potential fluctuation.

The inset in Fig.~\ref{f4}(a) shows $\Delta V$ as a function of temperature. $\Delta V$ is almost constant even as temperature and frequency change, indicating $E_{\mathrm{C}}$ and the capacitive coupling between the exit gate to QD are mostly constant \cite{holeJAP}. The difference in $\Delta V$ between the single-electron pumping and single-hole pumping would be because we used different QDs for each case.  

It is valuable to discuss a figure of merit parameter $\delta = \mathrm{ln}(\Gamma_{2}/\Gamma_{1})$, which is $(1+1/g)E_{\mathrm{C}}/(kT_{0})$ at low temperature and $(1+1/g)E_{\mathrm{C}}/(kT)$ at high temperature \cite{tunable-barrier1,holeJAP, GYgval}, where $g$ is the ratio between the potential change in QD and the entrance barrier height change with respect to the QD potential \cite{myPRB1}. $1/g$ appears due to the time difference between Figs.~\ref{f1}(e) and \ref{f1}(f). A large $\delta$ corresponds to a good pumping accuracy. This means that small $T_{0}$ (i.e., large $m$) and large $E_{\mathrm{C}}$ are important for high-accuracy pumping at low temperature. Since $\delta = A\Delta V$, we plotted $\delta$ as a function of temperature [Fig~\ref{f4}(b)]. Similar to Fig.~\ref{f4}(a), the light yellow and light blue eye guides are also phenomenologically assumed to have a function of $(T^{8} + T_{0}^{8})^{1/8}$ dependence. At low temperature, $\delta$ for the single-hole pumping are larger than those for the single-electron pumping, which is consistent with the estimated $\epsilon$ in Fig.~\ref{f3}. In addition, at more than 15 K, we observe crossover of $\delta$ between the single-electron pumping and single-hole pumping. This indicates that $(1+1/g)E_{\mathrm{C}}$ in the case of the single-electron pumping is larger than that in the case of the single-hole pumping, corresponding to the different $\Delta V$ between the two cases. These data indicate that a low $T_{0}$ in the case of the single-hole pumping is critical to achieve the high-accuracy pumping at low temperature even when $E_{\mathrm{C}}$ is smaller than that in the case of the single-electron pumping. Note that a large $E_{\mathrm{C}}$ is important for high temperature operation. We also note that the variation of $\delta$ is slightly larger in the case of the single-hole pumping, which is also seen in Fig.~\ref{f3}. This is related to the variation of $T_{0}$ discussed above and deserves further investigation.

In conclusion, we have compared the single-electron pumping and single-hole pumping using the same entrance barrier in a single ambipolar device. The single-hole pumping shows better characteristics in this device because of a low $T_{0}$ probably due to the heavy effective mass of holes. In future experiments, we will measure a device with the same QD and entrance barrier by fabricating a pump with four terminal contacts \cite{nob1} for a more systematic study.

We thank K. Chida for fruitful discussions.
%

\end{document}